\documentclass[superscriptaddress,aps,prb,nofootinbib,amsmath,amssymb,reprint]{revtex4-1}
\usepackage[utf8]{inputenc}
\usepackage{graphicx}
\usepackage{amsmath,amssymb}
\usepackage{textcomp}
\usepackage{color}
\usepackage{xcolor}
\usepackage{epstopdf}
\usepackage{siunitx}
\usepackage{mathtools}
\usepackage{natbib}
\usepackage[normalem]{ulem}
\usepackage{soul}

\newcommand{\vect}[1]{\boldsymbol{#1}}
\newcommand{\red}[1]{{\textcolor[rgb]{1,0,0}{#1}}}

\begin{document}

\title{Singlet state formation and its impact on magnetic structure in tetramer system SeCuO$_3$}

\author{Tonči Cvitani\'{c}}
\thanks{These two authors contributed equally}
\affiliation{Department of Physics, Faculty of Science, University of Zagreb, Bijeni\v {c}ka 32, Zagreb HR 10000, Croatia}
\author{Vinko \v{S}urija}
\thanks{These two authors contributed equally}
\affiliation{Institute of Physics, Bijeni\v {c}ka 46, HR-10000 Zagreb, Croatia}
\author{Krunoslav Pr\v{s}a}
\affiliation{Institute of Physics, EPFL, CH-1015 Lausanne, Switzerland}
\author{Oksana Zaharko}
\affiliation{Laboratory for Neutron Scattering and Imaging, PSI, CH-5232 Villigen, Switzerland}
\author{Peter Babkevich}
\affiliation{Institute of Physics, EPFL, CH-1015 Lausanne, Switzerland}
\author{Matthias Frontzek}
\affiliation{Laboratory for Neutron Scattering and Imaging, PSI, CH-5232 Villigen, Switzerland}
\author{Miroslav Po\v{z}ek}
\affiliation{Department of Physics, Faculty of Science, University of Zagreb, Bijeni\v {c}ka 32, Zagreb HR 10000, Croatia}
\author{Helmuth Berger}
\affiliation{Institute of Physics, EPFL, CH-1015 Lausanne, Switzerland}
\author{Arnaud Magrez}
\affiliation{Institute of Physics, EPFL, CH-1015 Lausanne, Switzerland}
\author{Henrik M. R\o{}nnow}
\affiliation{Institute of Physics, EPFL, CH-1015 Lausanne, Switzerland}
\author{Mihael S. Grbi\'{c}}
\email{mgrbic@phy.hr}
\affiliation{Department of Physics, Faculty of Science, University of Zagreb, Bijeni\v {c}ka 32, Zagreb HR 10000, Croatia}
\author{Ivica \v{Z}ivkovi\'{c}}
\email{ivica.zivkovic@epfl.ch}
\affiliation{Institute of Physics, Bijeni\v {c}ka 46, HR-10000 Zagreb, Croatia}
\affiliation{Institute of Physics, EPFL, CH-1015 Lausanne, Switzerland}

\begin{abstract}
We present an experimental investigation of the magnetic structure in a tetramer system SeCuO$_3$ using neutron diffraction and nuclear resonance techniques. We establish a non-collinear, commensurate antiferromagnetic ordering with a propagation vector $\textbf{k} = \left(0,0,1 \right)$. The order parameter follows a critical behavior near $T_N = 8$ K, with a critical exponent $\beta = 0.32$ in agreement with a 3D universality class. Evidence is presented that a singlet state starts to form on tetramers at temperatures as high as 200 K, and its signature is preserved within the ordered state through a strong renormalization of the ordered magnetic moment on two non-equivalent copper sites, $m_{\text{Cu1}} \approx \SI{0.4}{\mu_B}$ and $m_\text{Cu2} \approx \SI{0.7}{\mu_B}$ at 1.5 K.
\end{abstract}

\date{\today} 
\maketitle

\section{Introduction}

Classical, 3-dimensional magnetic systems tend to order in a long-range magnetic structure due to the presence of interaction $J$ between magnetic moments. Their behavior around the ordering temperature is well understood in terms of thermally induced fluctuations. More interesting behavior is found when the system is subject to quantum fluctuations, governed by an external parameter other than temperature. Quantum effects become especially important when the geometrical arrangement of magnetic moments induces frustration in selection of a unique ground-state via competing interactions and/or the effective dimensionality is reduced to $D < 3$. Recent investigations of triangular- and kagome-based systems~\cite{Li2016,Han2012} with signatures of a potential quantum spin-liquid phase reveal how quantum effects can completely prevent long-range order from occurring.

The ultimate limit of low-dimensional reduction is reached when a local cluster of spins is separated from other clusters around it. Significant attention has been devoted to single-molecule magnets, with strong single-ion anisotropy governing macroscopic quantum tunneling of magnetization~\cite{Sessoli1993}. In these metal-organic compounds the interaction between clusters is negligible due to large separation by organic ligands. On the other hand, in recent years several compounds have been discovered where the inter-cluster interaction allows the observation of subtle effects that govern the transition from local quantum states towards delocalized spin-waves. For instance in TlCuCl$_3$ pressure can be used as a tuning parameter for inter-cluster interaction, allowing the observation of closing the singlet-triplet gap and the emergence of long-range AFM order~\cite{Ruegg2004}. In other compounds, where the interactions are not easily modified, it remains an outstanding question of how the specific ratio of intra- to inter-cluster interaction(s) influences magnetic properties of a given compound. This becomes especially interesting when clusters are composed of more than just two AFM-coupled spins, increasing the number of local quantum states. With three spins in the cluster the ground state has a non-zero spin state and a divergent magnetic susceptibility when $T \rightarrow 0$. The 4-spin AFM-based clusters, called tetramers, again exhibit a singlet ground state but now the excited states comprise additional singlet state, three triplets and a quintet. Their relative order is determined by the geometry of the cluster and the relative strength of intra-cluster interactions~\cite{Haraldsen2005}.

Several compounds have been reported that adhere to the weakly-coupled tetramer model. A tetrahedron configuration of S=1/2 spins have been found in Cu$_2$Te$_2$O$_5$Cl$_2$ and Cu$_2$Te$_2$O$_5$Br$_2$, with intra-cluster interactions only twice larger than inter-cluster ones~\cite{Prester2004}. A diamond-shape cluster has been found in Cu$_2$PO$_4$OH, with intra-cluster interaction around 140 K and without long-range order down to 2 K~\cite{Kuo2008}. A linear tetramer system has been found in Cu$_2$CdB$_2$O$_6$~\cite{Hase2005}, SeCuO$_3$~\cite{Zivkovic2012} and CuInVO$_5$~\cite{Hase2016}, all showing strong intra-cluster interactions of 100-300 K, with long-range order occurring below 10 K, suggesting inter-cluster interactions two orders of magnitude smaller. 

\begin{figure}[h!]
	\centering
	\includegraphics[width=\columnwidth]{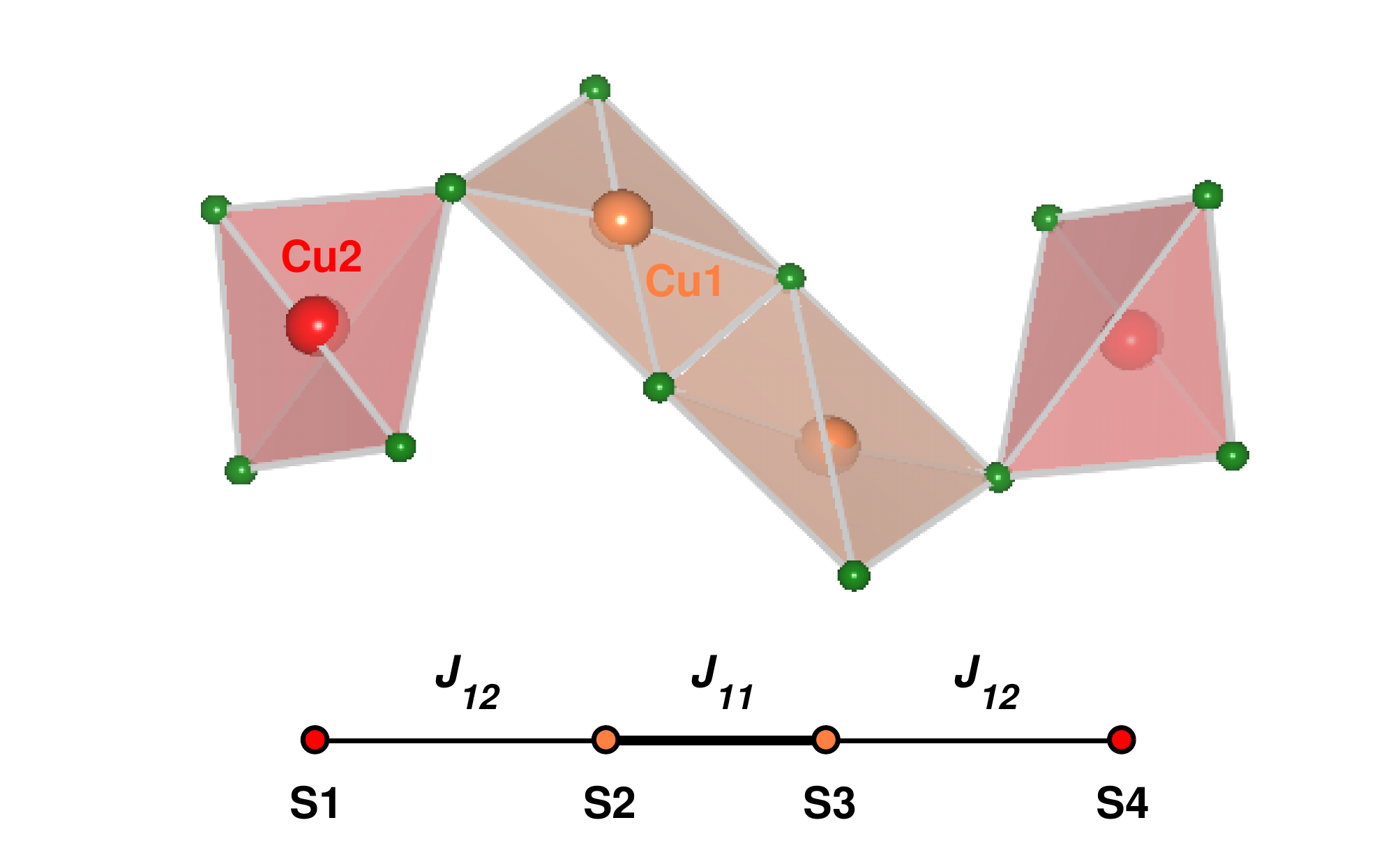}
	\caption{(Color online) A single Cu2 -- Cu1 -- Cu1 -- Cu2 tetramer, with shaded CuO$_4$ plaquettes. Below a schematic drawing of spins and exchange interactions between them (see Eq.~\ref{eq:tetra}).}
	\label{fig:scheme}
\end{figure}

In this article we focus our attention on SeCuO$_3$ and the peculiarities of its magnetic ground state. SeCuO$_3$ crystallizes in a monoclinic unit cell with the space group P2$_1$/$n$ as reported in Ref~\onlinecite{Effen}. It has two crystallographically inequivalent copper sites, Cu1 and Cu2. Each copper is surrounded by six oxygen atoms which form elongated octahedrons: the four nearest oxygen ions (distance around \SI{1.9}{\angstrom}) form CuO$_4$ \emph{plaquettes}, with apical oxygens further away (\SI{2.4}{\angstrom}). This crystal configuration places the $d_{x^2-y^2}$ -dominated orbital highest in energy, with a single magnetically active electron \cite{Zivkovic2012}. Each Cu1 site is connected via two oxygens to another Cu1 site (interaction $J_{11}$), and with a single oxygen to the Cu2 site (interaction $J_{12}$), effectively forming a Cu2 -- Cu1 -- Cu1 -- Cu2 linear tetramers (Fig. \ref{fig:scheme}). The Hamiltonian of a spin tetramer is expressed through the Heisenberg interactions:
\begin{equation}
\label{eq:tetra}
\mathcal{H} = J_{11} \mathbf{S_2 \cdot S_3} + J_{12} \left( \mathbf{S_1 \cdot S_2} + \mathbf{S_3 \cdot S_4} \right) \ .
\end{equation}

Previous results \cite{Zivkovic2012, Herak2014} showed that magnetic moments in SeCuO$_3$ form isolated tetramers, so the system behaves like a quasi zero-dimensional antiferromagnet. The tetramer Hamiltonian (Eq.~\ref{eq:tetra}) was used to explain experimental results down to $T \approx 90$~K, determining $J_{11}$ and $J_{12}$ to be approximately 225~K and 160~K, respectively. The system remains disordered down to $T_N = 8$~K where a 3D antiferromagnetic (AFM) order sets in. The magnetically ordered ground state emerges when the singlet ground state hybridizes with the excited states due to mixing by intertetramer interactions and/or anisotropy terms. No clear model exists to explain the experimental results in the intermediate phase (8 K -- 90 K).  The tetramer model is not able to reproduce the steep increase in magnetic susceptibility but a broad maximum around 20 K suggests low-dimensional behavior. An interesting aspect has been revealed by ESR and torque magnetometry~\cite{Herak2014} where a strong rotation of magnetic axes in the  intermediate temperature range  was observed. 
Recent magnetization measurements at high magnetic field, ESR and $^{77}$Se NMR reported a magnetic anomaly in the ordered phase below $T = 6$~K which supposedly occurs due to a spin reorientation (Ref.~\onlinecite{Lee}).

In this paper we address the aspects of the magnetic structure within the ordered state using neutron diffraction and $^{63,65}$Cu NQR resonance measurements. We found no trace of the anomaly or spin reorientation reported in Ref.~\onlinecite{Lee} and present evidence of the Cu1-Cu1 spin singlet formation below $T \approx 200$~K.

\section{Experimental details}

Single crystals of SeCuO$_\text{3}$ were grown by a standard chemical vapor phase method. Mixture of analytical grade purity CuO and SeO$_\text{2}$ powders in molar ratio 4:3 was sealed in the quartz tubes with electronic grade HCl as the transport gas for the crystal growth. The ampules were then placed horizontally into a tubular two-zone furnace and heated very slowly by \SI{50}{\textdegree C/h} to \SI{500}{\textdegree C}. The optimum temperatures at the source and deposition zones for the growth of single crystals were \SI{550}{\textdegree C} and \SI{450}{\textdegree C}, respectively, and after four weeks many green SeCuO$_3$ crystals were obtained. The phase purity was verified using x-ray powder diffraction. 

Neutron powder diffraction patterns were collected on the Cold Neutron Powder Diffractometer (DMC) at SINQ, Paul Scherrer Institute, Switzerland. The 4.2 g SeCuO$_\text{3}$ sample was sealed in a 8 mm diameter cylindrical vanadium can under helium and cooled down in the Orange ILL type cryostat. Neutron wavelength of $\lambda=\SI{2.46}{\angstrom}$ was used and about \SI{6}{h} was needed for one temperature point within the 1.5 K-15 K range.

Single crystal neutron diffraction was performed on thermal single crystal neutron diffractometer (TriCS) at SINQ, PSI, Switzerland. A large single crystal (\SI[product-units=power]{17x5x5}{mm}) was mounted on an Al "T"-shaped sample holder using Al wire and placed inside an ILL Orange type cryostat. Diffraction was measured in the temperature range from \SI{1.5}{K} to \SI{15}{K} using a wavelength of $\lambda=\SI{2.317}{\angstrom}$ in the normal beam geometry. Both the powder and single crystal data were analyzed and refined using the FullProf suite\cite{FullProf}.

To get a better insight into  the local spin behavior, nuclear magnetic resonance (NMR) and quadrupolar resonance (NQR) techniques were used. These techniques locally probe interactions with an observed nuclei, being sensitive to external magnetic field and to electric field gradient (EFG) on the nuclei. Since the magnetically active electrons reside on the Cu sites, $^{63,65}$Cu NMR/NQR was chosen for this experiment, as it provides direct insight into the local spin behavior.

The NMR/NQR experiment was conducted in a superconducting Oxford Instruments variable-field magnet. The sample was a single-crystal (\SI[product-units=power]{5x1x1}{mm}), with  the longest axis along the $b$-axis. A copper coil was tightly wound around the longest axis of the sample, which was mounted on a single axis rotator with the axis of rotation within $5^\circ$ from the $b$-axis. Spectra were acquired with a Tecmag Apollo spectrometer by the standard Hahn echo sequence $\pi/2 - \tau - \pi$, followed by Fourier transform of the echo signal. The duration of the $\pi/2$ pulse was 1.4~$\mu$s, and the time between the pulses was varied from $\tau = 15$~$\mu$s at low temperatures to $\tau = 3$~$\mu$s at $T = 200$~K and above. Multiple spectra were added together to obtain a broadband spectra by the usual variable offset cumulative spectroscopy method.

\section{Experimental results}

\subsection{Neutron powder diffraction}

\begin{figure}
	\centering
	\includegraphics[width=\columnwidth]{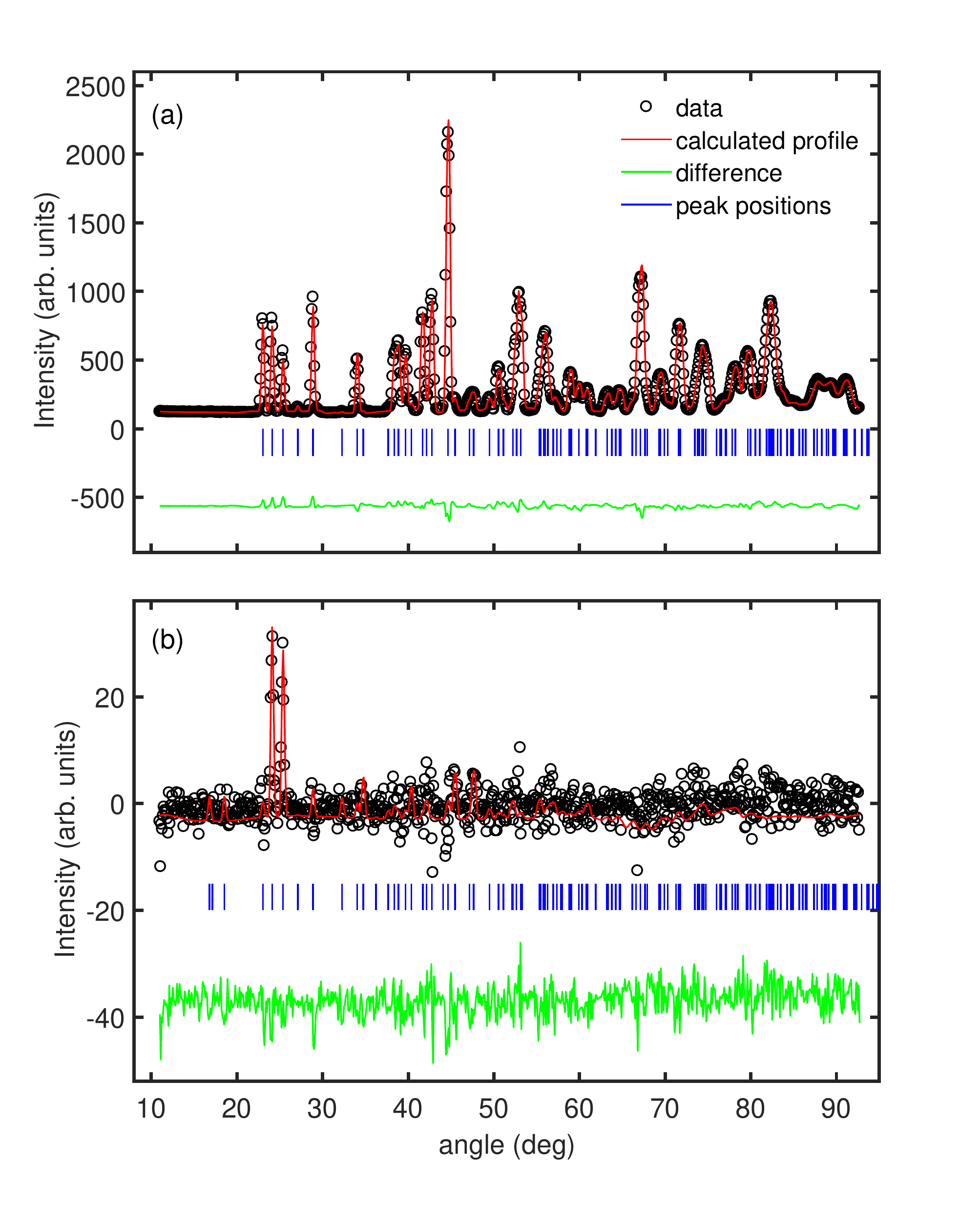}
	\caption{(Color online) (a) Neutron powder diffraction pattern of SeCuO$_\text{3}$ at \SI{15}{K}. The solid red line is calculated pattern for SeCuO$_\text{3}$ using the space group P2$_1$/$n$, while the solid green line indicates the difference between calculated and observed intensity. Vertical bars indicate calculated positions of Bragg reflections. (b) Magnetic diffraction pattern at \SI{1.5}{K} after subtracting pattern at \SI{15}{K}. Solid red line represents a calculated intensity using the $\Gamma_4$ IR.}
	\label{fig:npd_HT}
\end{figure}

Fig.~\ref{fig:npd_HT}(a) shows the powder neutron diffraction pattern of SeCuO$_\text{3}$ measured at \SI{15}{K}. Refinement was performed against the model presented in Ref.~\onlinecite{Zivkovic2012}. The agreement between the published and refined unit cell parameters, a=\SI{7.6989 \pm 0.0002}{\angstrom}, b=\SI{8.2126 \pm 0.0002}{\angstrom}, c=\SI{8.4856 \pm 0.0002}{\angstrom} and $\beta$=\SI{99.178 \pm 0.002}{\textdegree}, is very good.

The difference between the patterns collected at \SI{15}{K} and \SI{1.5}{K} is shown in Fig.~\ref{fig:npd_HT}(b). The difference pattern contains only two magnetic reflections, since the structural changes between the two temperatures are negligible. The magnetic peaks are found on top of nuclear ones making separation of the purely magnetic signal challenging.

Using the K-SEARCH program from the FullProf suite we determined the magnetic propagation vector $\textbf{k} = \left(0,0,1 \right)$. This reveals commensurate antiferromagnetic ordering. To constrain the possible magnetic structures, we have performed irreducible representation analysis using the program BASIREPS from the FullProf suite. The symmetry analysis for the propagation vector $\textbf{k} = \left(0,0,1 \right)$ and space group P2$_1$/$n$ yielded only four nonzero one-dimensional $\left( \Gamma^1 \right)$ irreducible representations (IRs), each with the multiplicity of 3. The magnetic representation $\Gamma_\text{mag}$ is thus composed of:
\begin{equation}
\Gamma_\text{mag} = 3 \Gamma_1 + 3 \Gamma_2 + 3 \Gamma_3 + 3 \Gamma_4,
\end{equation}
where superscript 1 is omitted for clarity. The basis vectors for these IRs are listed in Table~\ref{tab:vects}.

\begin{table}
\centering
\caption{Basis functions of the irreducible representations for the Cu1 and Cu2 sites (using the same notation as in Ref.~\onlinecite{Zivkovic2012}) in the space group P2$_1$/$n$ with the propagation vector $\textbf{k} = \left(0,0,1 \right)$ for SeCuO$_\text{3}$ obtained from the representational analysis. Each crystallographic Cu site has four Cu atoms: Cu$_1$: $\left(x,y,z\right)$, Cu$_2$: $\left(-x+1/2, y+1/2 ,-z+1/2\right)$, Cu$_3$: $\left(-x,-y,-z\right)$ and Cu$_4$: $\left(x+1/2, -y+1/2,z+1/2\right)$. Since all basis functions of these IRs contain only real components, and all have only diagonal elements, only the diagonal terms are given in table below. \label{tbl:IRs}}
\label{tab:vects}
\begin{tabular}{llrrrr}
\hline
\hline
IRs & Site: & Cu$_1$ & Cu$_2$ & Cu$_3$ & Cu$_4$    \\
\hline
$\Gamma_1$ & & $\left( 1,  1,  1\right)$ & $\left(-1,  1, -1\right)$ & $\left( 1,  1,  1\right)$ & $\left(-1,  1, -1\right)$ \\ 
$\Gamma_2$ & & $\left( 1,  1,  1\right)$ & $\left(-1,  1, -1\right)$ & $\left(-1, -1, -1\right)$ & $\left( 1, -1,  1\right)$ \\
$\Gamma_3$ & & $\left( 1,  1,  1\right)$ & $\left( 1, -1,  1\right)$ & $\left( 1,  1,  1\right)$ & $\left( 1, -1,  1\right)$ \\
$\Gamma_4$ & & $\left( 1,  1,  1\right)$ & $\left( 1, -1,  1\right)$ & $\left(-1, -1, -1\right)$ & $\left(-1,  1, -1\right)$ \\
\hline
\end{tabular}
\end{table}

\begin{figure}[h!]
	\centering
	\includegraphics[width=\columnwidth]{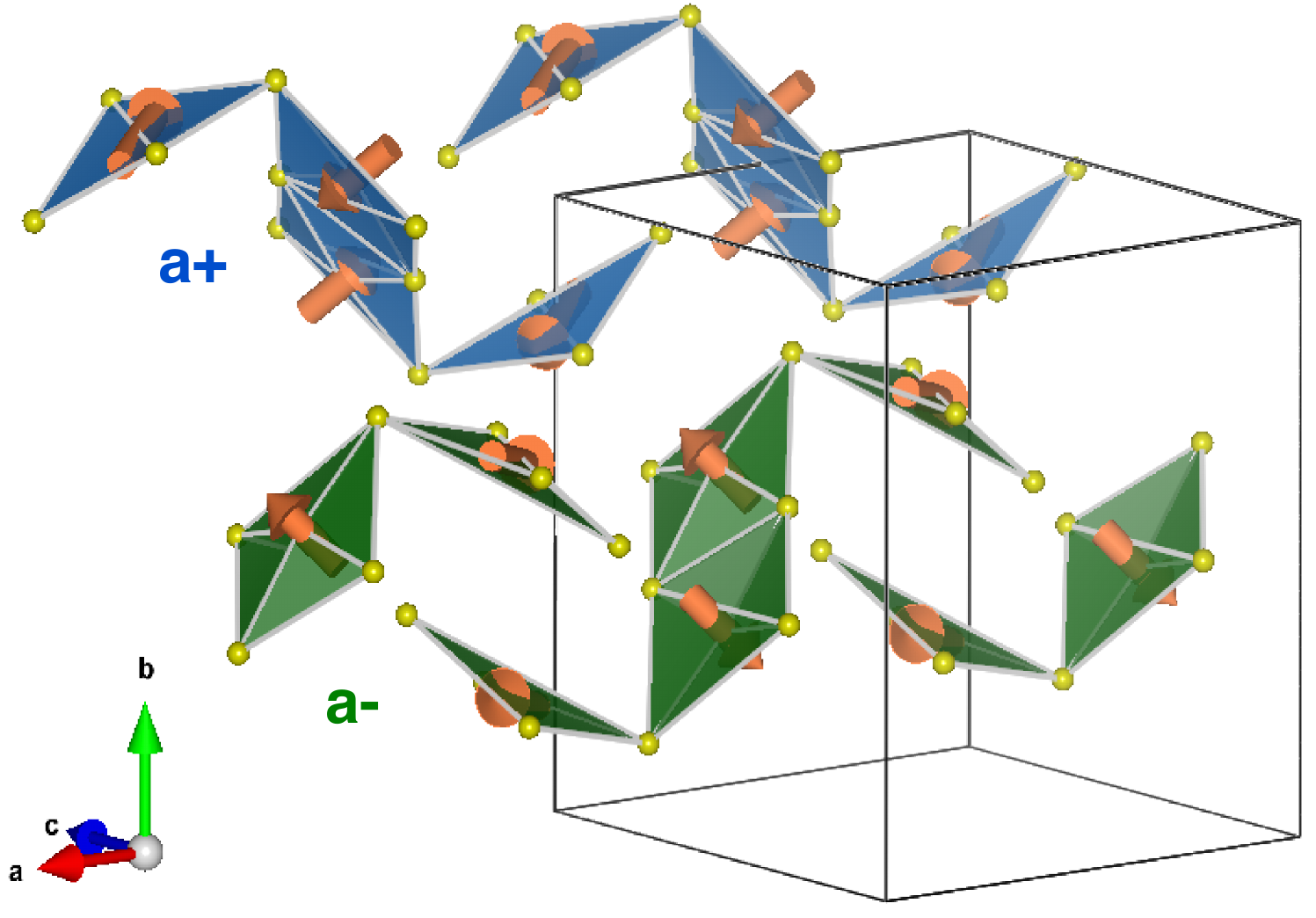}
	\caption{(Color online) Magnetic structure of SeCuO$_\text{3}$. Two chains of tetramers (blue and green) are oriented along the $a$-axis, with a different sense of direction ($a+$ and $a-$, respectively). They form inter-penetrating lattices, with green chains propagating in the middle of the unit cell, while the blue ones - along its edges. Only one blue chain is shown for clarity. Selenium atoms are omitted.\cite{Momma:db5098}}
	\label{fig:mag_tetramers}
\end{figure}

Two solutions: 1) $\Gamma_4$ for the Cu1 site and $\Gamma_2$ for the Cu2 site and 2) $\Gamma_4$ IR for both sites, gave equally good agreement with the data. As the transition from the paramagnetic state to the magnetically ordered state is of the second order, we conclude that both copper sites must order with the same $\Gamma_4$ IR.

The best agreement for both, powder and single crystal diffractions, is found for the moment configuration presented in Fig.~\ref{fig:mag_tetramers}. As shown in Ref.~\onlinecite{Zivkovic2012}, the active orbitals for the magnetic exchange in SeCuO$_3$ are $d_{x^2-y^2}$. Accordingly, the structure can be represented as a network of weakly coupled tetramers. Each tetramer has a shape of a 'sea-horse', forming a one-dimensional chain of tetramers along the $a$-axis with a 'head-over-tail' coupling between the neighbours (weak inter-tetramer $J_{22}$ interaction between two Cu2 moments mediated by an almost perpendicular Cu -- O -- Cu bridge). There are two chains, one running along the '$a+$' direction and one along the '$a-$' direction, in Fig.~\ref{fig:mag_tetramers} these chains are shown by blue and green, respectively. Tetramers from different chains are symmetry related by the 180$^{\circ}$ rotation, followed by the $[ \,100] \,$ translation. The coupling across the chains is mediated through the network of SeO$_3$ tetrahedra (not shown).

Several features of the magnetic structure presented in Fig.~\ref{fig:mag_tetramers} are worth emphasizing. We see that an anti-parallel arrangement of spins is formed on the equivalent copper sites, $\uparrow_{Cu1}$ -- $\downarrow_{Cu1}$ and $\uparrow_{Cu2}$ -- $\downarrow_{Cu2}$. On the other hand, spins on the neighbouring Cu1 and Cu2 ions have an angle of $\approx 97^{\circ}$, indicative of a significant Dzyaloshinskii-Moriya interaction due to the lack of the inversion center between the Cu1 and Cu2 ions. Additionally, due to the rotational symmetry between the '$a+$' and '$a-$' chains (see Fig.~\ref{fig:mag_tetramers}), the overall magnetic structure is highly non-collinear.

Among the most important results of our analysis is the magnitude of magnetic moments on the copper ions. They amount to $m_{\text{Cu1}} = \SI{0.46}{\mu_B}$ and $m_\text{Cu2} = \SI{0.73}{\mu_B}$, indicating a strong renormalization from the nominal $\SI{1}{\mu_B}$ value. The fact that the inner copper sites (Cu1) have smaller ordered moments than the outer ones (Cu2) supports the initial model based from the susceptibility analysis~\cite{Zivkovic2012,Herak2014}, where the $J_{11}$ coupling is notably stronger than $J_{12}$ . This induces stronger quantum fluctuations from the singlet state.

\subsection{Nuclear quadrupolar resonance -- paramagnetic phase}

\begin{figure}[h!]
\centering
\includegraphics[width = 1\linewidth]{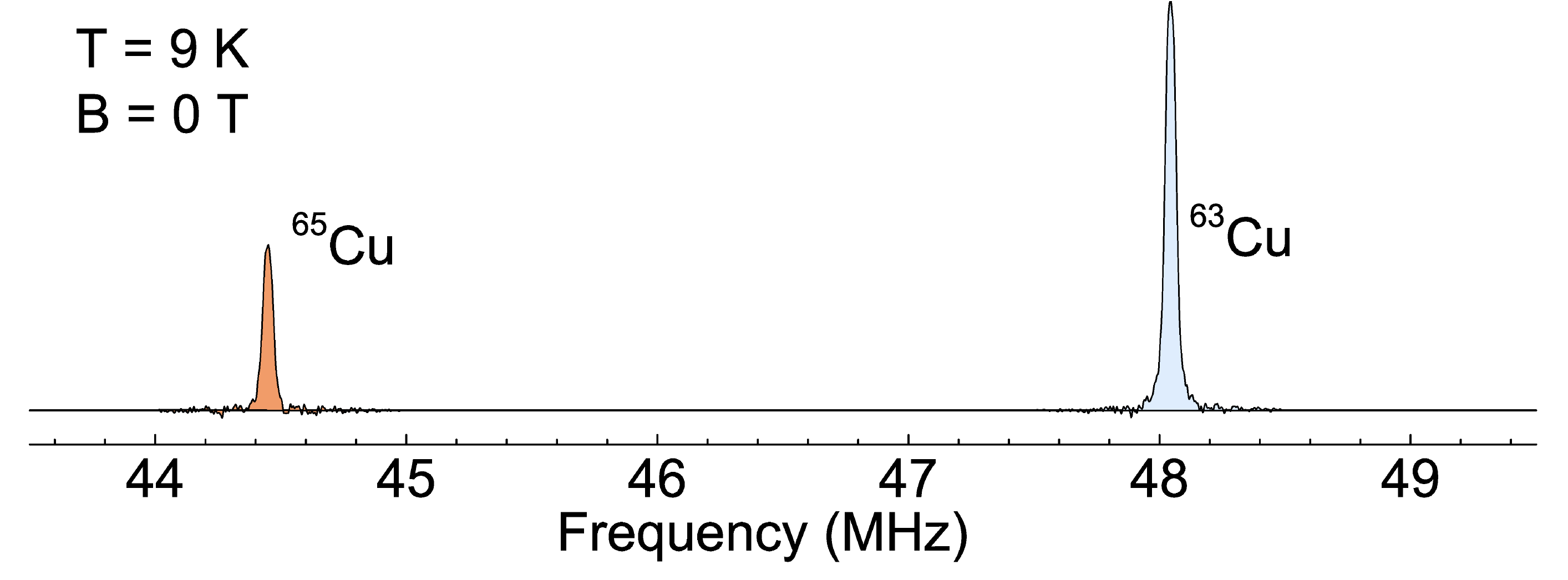}
\caption{(Color online) NQR spectrum at $T = 9$~K, above antiferromagnetic transition. Two different Cu isotopes are labelled and coloured accordingly: red for $^{65}$Cu and blue for $^{63}$Cu. From crystal structure, we should see two copper sites (Cu1 and Cu2), so the number of lines should be four (two for each isotope). Missing lines imply that one copper site has fast relaxation rate, and it is not visible by NQR/NMR.}
\label{fig:PMlines}
\end{figure}

The NMR and NQR methods probe nuclear spins and their coupling to the local magnetic field and EFG, respectively. EFG is a tensor defined as $V_{ij} = \partial V / \partial x_i \partial x_j$, usually expressed with principal axes following the standard convention $|V_{zz}| > |V_{yy}| > |V_{xx}|$. Within SeCuO$_3$, copper nuclei have nonvanishing EFG that give rise to an NQR signal. The local symmetry of the copper sites implies that $V_{zz}$ is oriented perpendicular to the plane defined by the 4 oxygens in each CuO$_4$ plaquette.

Above $T_N$ the NQR spectra show two distinct lines that correspond to two copper isotopes, $^{65}$Cu and $^{63}$Cu, shown in Fig.~$\ref{fig:PMlines}$. The lines can be fit to a Gaussian function, with FWHM of 40~kHz. The line width is constant in the temperature range from $T = 9$ to $T = 260$~K, indicating no onset of short range order. 

With two inequivalent crystallographic sites (Cu1 and Cu2), one would expect different EFG values for each of them, so that four NQR lines should be visible, two for each isotope. A second pair of lines was searched for in a broad frequency range, from 16 to 60~MHz. Lack of the second pair of lines, even with short interpulse time ($\tau = 3 \ \mu$s), implies that one copper site has too fast relaxation rate to be detected by NQR.

\begin{figure}[h!]
\centering
\includegraphics[width = 1\linewidth]{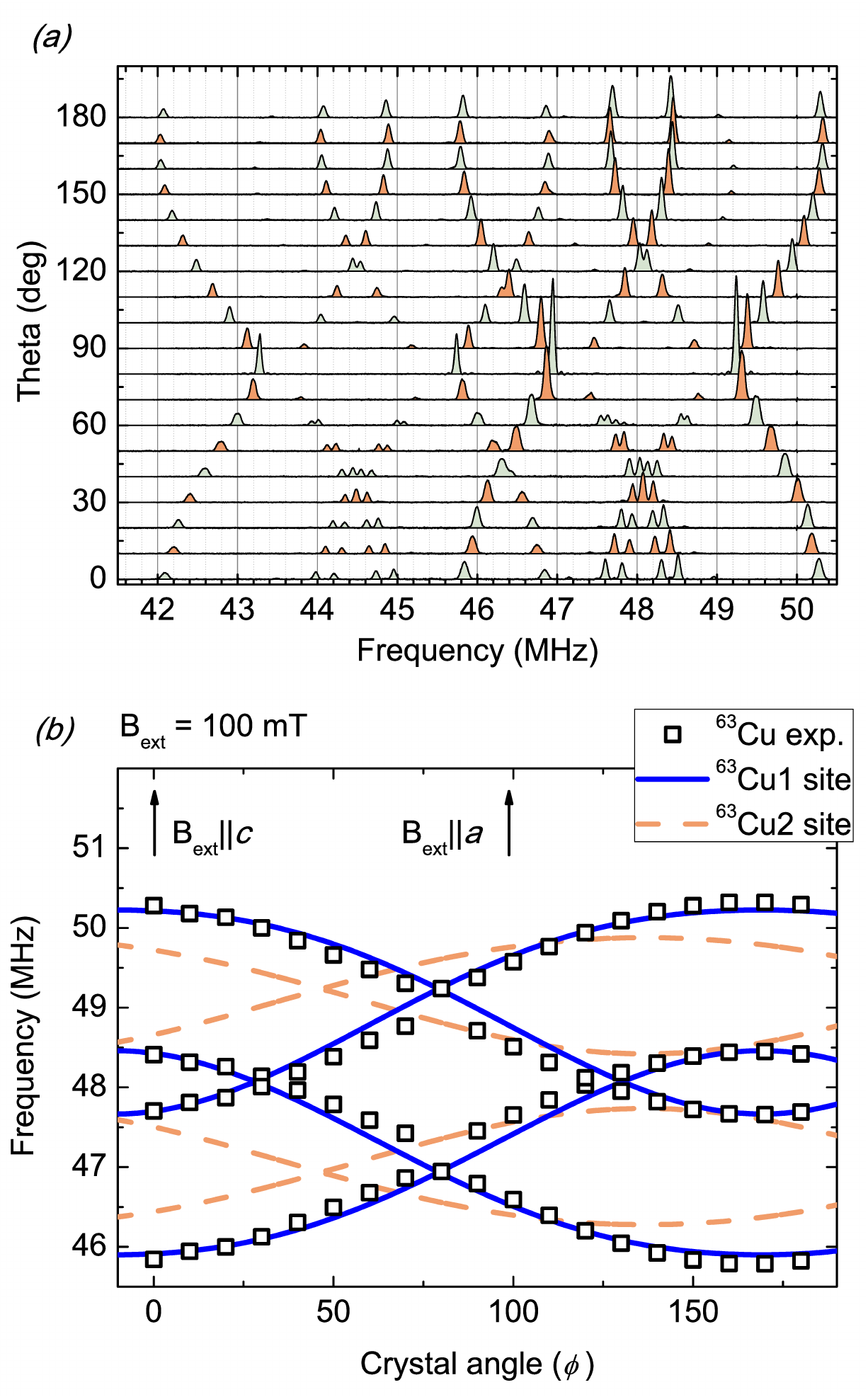}
\caption{(Color online)\textit{(a)} Rotational field perturbed NQR spectra. $B = 100$~mT, $T = 19$~K. Eight lines can be easily distinguished (four per each isotope). Small additional line splitting (near $0^\circ$) arises from small misalignment: b-axis of the sample is not completely perpendicular to external magnetic field. \textit{(b)} Line position of $^{63}$Cu at different sample orientation (white squares) from \textit{(a)}, $T = 19$~K. Crystal angle is the angle between the crystallographic $c$-axis and external magnetic field ($B_{ext} = 100$~mT). Simulation of line position (see text) for two copper sites is shown: Cu1 (solid blue line) and Cu2 site (dashed green lines). Cu1 site better fits observed data, since Cu2 site lacks crossing of inner lines (at $30^\circ$ and $120^\circ$). Only two lines are present at $80^\circ$, implying asymmetry parameter $\eta = 0$ (see text).}
\label{fig:rotSpec}
\end{figure}

To determine which Cu site is observed by NQR, we measured an angle dependence of Zeeman perturbed NQR of SeCuO$_3$ (Fig. \ref{fig:rotSpec} (a)). By applying an external field the single NQR line splits into four lines, whose positions depend on the angle of the external magnetic field with respect to the principal axis $V_{zz}$. The sample rotation was performed around the $b$-axis, while the magnetic field was applied within the $ac$-plane.

The complete NMR/NQR Hamiltonian~\cite{slichter} is:
\begin{equation}
\mathcal{H} = -\gamma_n \hbar B I_z \cos \vartheta + \frac{e Q V_{zz}}{4 I (2I - 1)} \left( 3 I_z^2 - I^2 + \eta (I_x^2 - I_y^2) \right) \ ,
\label{eq:ham}
\end{equation}
where $\eta = (V_{xx}-V_{yy})/V_{zz}$ is the asymmetry parameter, $\vartheta$ is the angle between  the local $z$-axis and the local magnetic field $\vect{B}$, $\mathbf{I}$ is the nuclear spin (for copper $I = 3/2$), and $Q$ and $\gamma_n$ are the quadrupole moment and the gyromagnetic ratio of the nucleus, respectively. Numerical diagonalization of the Hamiltonian was used to explain the spectra. Four lines instead of two arise due to mixing of nuclear spin states that is beyond first-order perturbation theory.

Free parameters are the asymmetry $\eta$ and the angle $\vartheta$. $V_{zz}$ and $\eta$ are interdependent, and their relation can be deduced from NQR at $B = 0$~T (Fig.~\ref{fig:PMlines}). In that case, the line position is given as $\nu_Q = \frac{e Q V_{zz}}{2} \sqrt{1 + \eta^2 / 3}$. 
Since two crystallographic sites (Cu1 and Cu2) will have different $\vartheta$ angles, consequentially they will have different line positions which allows us to determine which site is visible in the experiment.

The calculated line positions for $B=B_{ext}=100$~mT and experimental results for rotation spectra around the crystal $b$ axis are shown in (Fig.~\ref{fig:rotSpec}(b)). Two distinct features are clearly visible. The crossing of inner lines at angles 30 and 120 implies $\vartheta$ to be equal to the magic angle~\cite{magic}. Analysing the crystallographic data and the direction of rotation we find that only the NQR signal of the Cu1 site can reproduce this crossing.Therefore, the observed signal comes from the Cu1 site. 
Other feature of the NQR spectrum is the existence of a single line at $80^\circ$, which implies $\eta = 0$. Thus, all parameters of the Hamiltonian are determined.

\subsection{Nuclear quadrupolar resonance -- AFM phase}

\begin{figure}[h!]
\centering
\includegraphics[width = 1\linewidth]{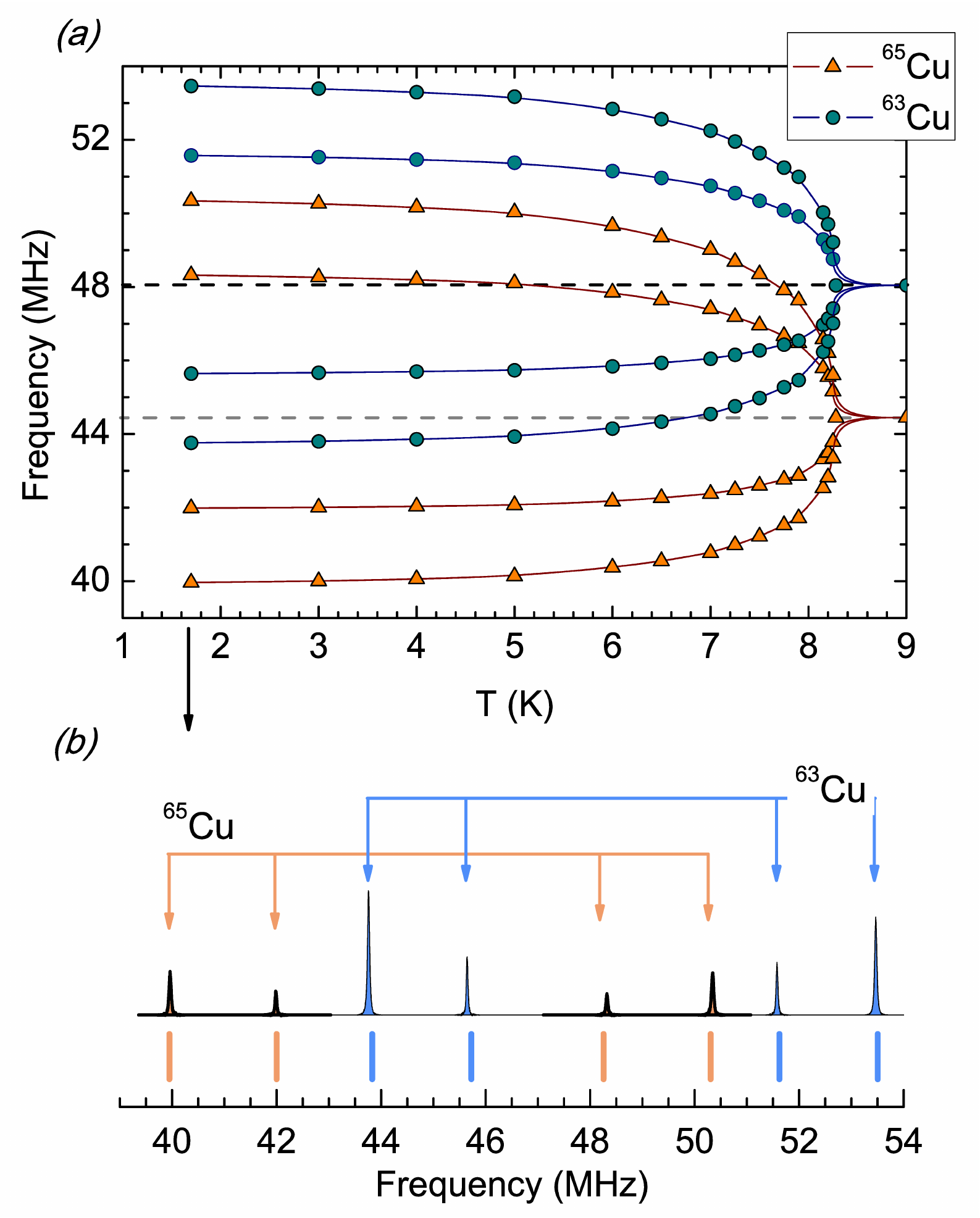}
\caption{(Color online) \textit{(a)} NQR line positions with respect to temperature. Lines are split because of staggered magnetic field $B_{hf}$. As temperature is lowered, $B_{hf}$ increases monotonously, without rotation of vector in crystal axes (see text). \textit{(b)} Line positions of obtained spectrum at $T = 1.7$~K (as indicated by arrow), with simulation of line positions denoted by vertical bars under the spectrum. Parameters are $B_{hf} = 0.35$~T, $\vartheta_B = 80.7^\circ$.}
\label{fig:Cu1spread}
\end{figure}

Below temperature $T_N$, the system enters a magnetically ordered state and each NQR line splits into four lines without external field, indicating the existence of the local magnetic field (staggered antiferromagnetic field). In Fig.~\ref{fig:Cu1spread}(a) we plot the line positions with respect to temperature. By lowering temperature  the staggered field intensity increases thus  shifting the lines further apart. This behavior is still governed by the same Hamiltonian (Equation \ref{eq:ham}), but the magnetic field $B$ and its angle $\vartheta$ are now governed by spin interactions. We can determine the line position for each temperature, with free parameters $B = B_{hf}, \vartheta = \vartheta_B$ and $\nu_Q \propto V_{zz}$. Latter parameter is expected to remain constant since it depends only on the crystal electric field $V_{zz}$, and neutron scattering shows  negligible variation of the crystal parameters in this temperature region.

Fig.~\ref{fig:Cu1spread}(b) presents the NQR fit for the lowest temperature, where the local magnetization is completely developed. Here we obtain $B_{hf} ( T=1.7 K) = 0.35$~T and $\theta_B = 80.7^\circ$. To fit the data at other temperatures it was sufficient to vary only the intensity of magnetic field $B_{hf}$, while the angle $\theta_B = 80.7^\circ$ remained constant. Thus, we can determine that there is no rotation of the magnetic moments  in the ordered state within the sensitivity of our measurements. This is consistent with the previous ESR results which found only a minor rotation (close to $T_N$) of the magnetic axes.\cite{Herak2014}

We can relate the magnetization vector $\boldsymbol{m}$ on the Cu1 site with hyperfine field:
\begin{equation}
\label{eq:hf}
\vect B_{hf} = A_{hf} \cdot \vect m
\end{equation}
where $A_{hf}$ is the hyperfine tensor. Since the hyperfine tensor is governed by the same crystal symmetries as the EFG tensor, it is justified to presume that their principal axes are the same. We take the hyperfine tensor $A_{zz} = -10$~T/$\mu_B$, $A_{xx} = A_{yy} = -1$~T/$\mu_B$. This result is from Ref.~\onlinecite{Itoh95} on compound CuGeO$_3$, but similar hyperfine tensor can be found in CuO and other insulators that have a Cu$^{2+}$ ion in an octahedral environment (e.g. in azurite~\cite{Aimo2009}). Inverting equation~\ref{eq:hf} and using data obtained from the NQR experiments, we can deduce the value of the magnetic moment residing on the Cu1 site to be $m = 0.35$~$\mu_B$, which is slightly smaller than moment extracted from neutron measurements ($m = 0.46$~$\mu_B$). This is to be expected, since the time scale of neutron measurements is much shorter than the NQR time scale and a larger influence of quantum fluctuations can be expected.

Additionally, analysis suggests that the spin on the Cu1 site lies almost completely inside the CuO$_4$ plaquette with $\vartheta_m = 89^\circ$. This differs from the value deduced from neutron measurements (Fig. \ref{fig:mag_tetramers}), where the Cu1 moments are tilt approximately by $45^\circ$ from the plaquette. Since the magnetic neutron intensity is relatively weak, the uncertainty of the direction of moment(s) is relatively large. Thus, the determined orientation of the moment on Cu1 should be more accurate from the NQR measurements. It is worth mentioning that, if the Cu1 magnetic moment direction is fixed to lie inside the plaquette, the resulting fit is noticeably worse than for the configuration presented in Fig.~\ref{fig:mag_tetramers}. This small controversy might be resolved with a future polarized neutron scattering experiment.

\begin{figure}
\centering
\includegraphics[width = 1\linewidth]{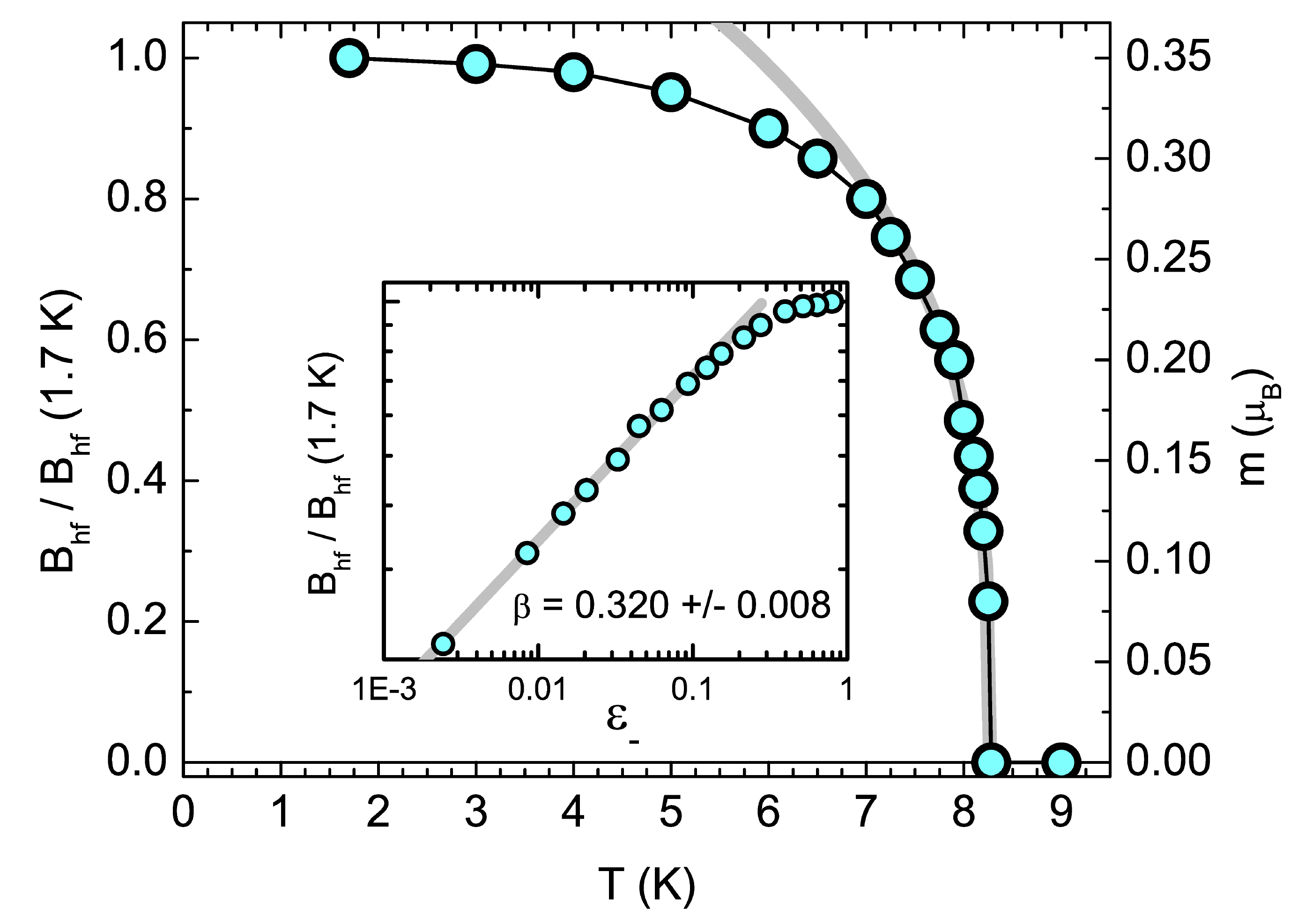}
\caption{(Color online) Staggered field as the order parameter extracted from spectra. Gray line is the fit to expression \ref{eq:ord}. $B_{hf} (1.7\mbox{~K}) = 0.35$~T. Scale on the right shows corresponding value of magnetization calculated from approximated hyperfine interaction. It is in agreement with neutron measurements (see text). Inset: Logarithmic plot shows power law behavior.}
\label{fig:orderparam}
\end{figure}

Since the staggered field $B_{hf}$ is proportional to the order parameter of the AFM phase transition (Fig.~\ref{fig:orderparam}){\red {,}} we can deduce a critical behavior below the transition ($T < T_N$) as follows:
\begin{equation}
\label{eq:ord}
\frac{B_{hf}(T)}{B_{hf}(0)} = \left( \frac{T_N - T}{T_N} \right)^\beta = \varepsilon_-^\beta
\end{equation}
The critical exponent was extracted from a fit in the range $\varepsilon_- = 2 \cdot 10^{-3} - 2 \cdot 10^{-1}$ (inset, Fig.~\ref{fig:orderparam}). The fitted value is shown in Table~\ref{tbl:crit}, together with predictions for several other universality classes. It is obvious that the extracted critical exponent is consistent with the 3D phase transition. A similar $\beta$-value was found by Lee \emph{et al.} in Ref.~\onlinecite{Lee}.

\begin{table}[h!]
\centering
\caption{Critical exponent $\beta$ extracted from temperature dependence of the order parameter below $T_N$. Listed are values of $\beta$ for universality classes most suited for this problem. \label{tbl:crit}}
\label{tab:ads}
\begin{tabular}{l c||c|c|c|c}
\hline
 & SeCuO$_3$ & 2D-Ising & 3D-Ising & 3D-Heisenberg & Mean field\\
$\beta$ & 0.32 & 0.13 & 0.33 & 0.35 & 1/2 \\
\hline
\end{tabular}
\end{table}

\subsection{Cu1 spin-lattice relaxation}
\begin{figure}[h!]
	\centering
	\includegraphics[width=\columnwidth]{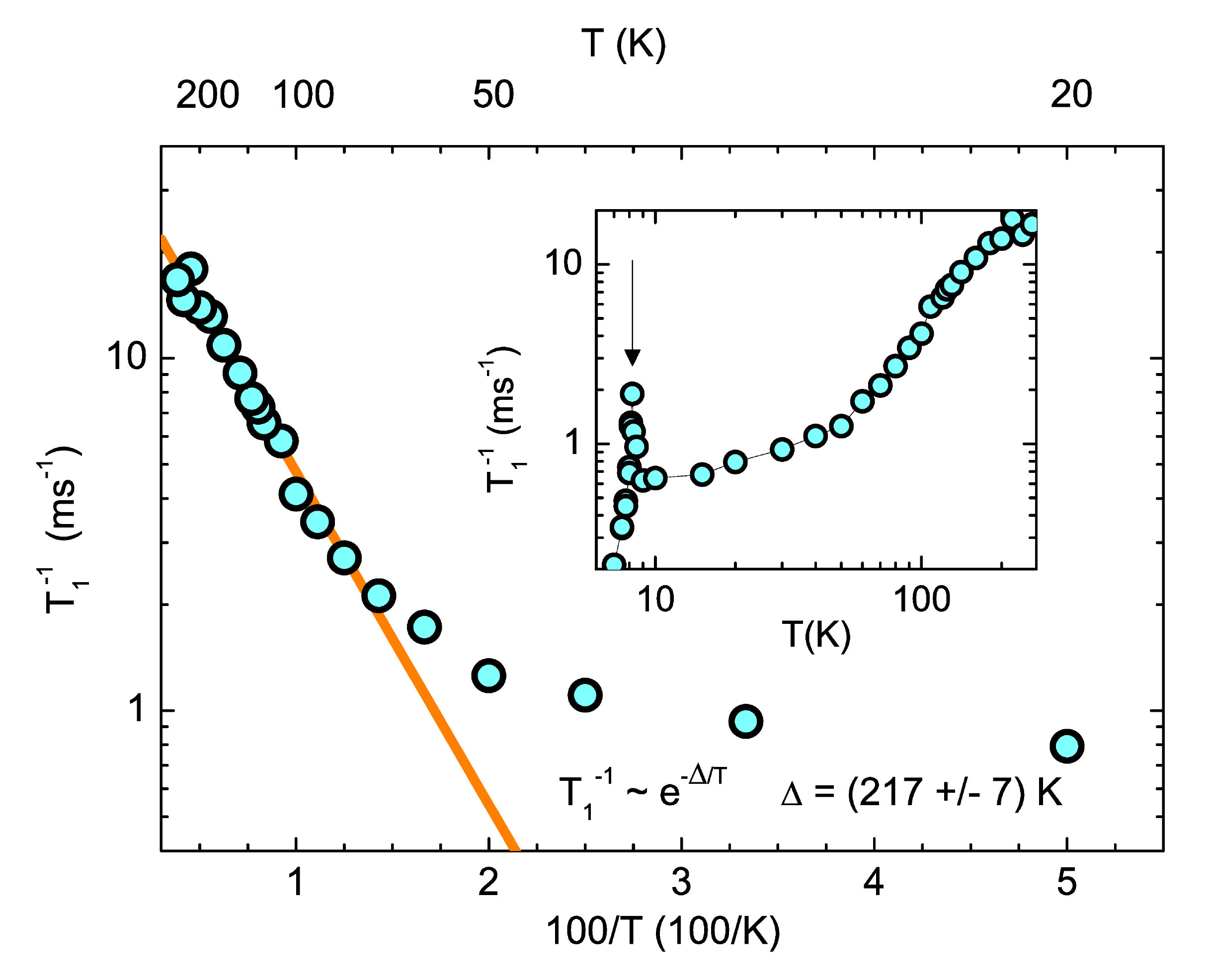}
	\caption{(Color online) Arrhenius plot of the Spin-lattice relaxation rate: note the inverse temperature and semilog plot. Points from 90--200~K were fitted. Sharp rise in relaxation with respect to temperature rise is attributed to breaking the Cu1 -- Cu1 spin singlet. Inset: Log-log scale. Arrow indicates critical fluctuations at $T_N = 8$~K. Above 200~K spin singlet is destroyed which results in small signal (hence large error bars) and constant relaxation rate (within experimental error).}
	\label{fig:tetraGap}
\end{figure}

To get a better insight into the spin dynamics above ordering temperature, the spin-lattice relaxation rate ($1/T_1$) was recorded with saturation recovery method and is shown in Fig. \ref{fig:tetraGap}. What can be clearly seen is the activated behavior. Relaxation increases by an order of magnitude from $T = 90$~K to $T = 200$~K above which the temperature dependence saturates and $T_1$ remains almost constant. This is expected in the paramagnetic state of antiferromagnetic insulators~\cite{Moriya56}, and shows that the truly paramagnetic state exists only at high temperatures. From the Arrhenius plot we can fit the activation energy of $\Delta = 217 \pm 7$~K, which is close to $J_{11}$ value determined from the susceptibilty.\cite{Zivkovic2012} Large error bars in relaxation at high temperatures are caused by the loss of signal due to shortening of spin-spin relaxation time ($T_2$). The activation behaviour indicates that an energy gap opens in the system. Since the ground state of an isolated tetramer system is singlet, the observed temperature dependence most likely originates from a singlet-triplet gap, which means that below 200~K singlets form at the Cu1 sites (dimer), as was proposed in an earlier work.\cite{Zivkovic2012} At temperatures lower than 70~K, additional parts of the isolated-tetramer Hamiltonian start to contribute to the relaxation process, and the  $1/T_1 (T)$ dependence deviates from activated.




\section{Discussion}
In earlier papers \cite{Zivkovic2012, Herak2014} it was suggested that the unusual behavior of the spin susceptibility arises from the rotation of the macroscopic magnetic axes as the temperature is lowered. This was additionally supported by the EPR and torque measurements, which mapped the temperature dependence of the magnetic axis shift.\cite{Herak2014} The authors hypothesized that this behavior is related to the formation of a singlet state on the inner Cu1 -- Cu1 pair, where the contribution from the Cu1 site's local axes of anisotropy to the total magnetic anisotropy diminishes and the net magnetization axes rotate towards those defined by the Cu2 site.

The evidence accumulated from two microscopic techniques presented in this article strongly supports this hypothesis. Additional possibilities included a dynamic Jahn-Teller effect, or a structural transition but the smooth temperature dependence of the NQR line (not shown) excludes this possibility. At high temperatures the $T_1$ relaxation data measured at the Cu1 site show a clear exponential decay that indicates a singlet-triplet gap. On the other hand, below $T_N$, when the system enters an ordered state driven by inter-tetramer interactions, both neutron diffraction and NQR indicate greatly reduced value of the ordered magnetic moment residing on the Cu1 site ($\approx 40$\% of fully polarized spin even at $T  = 1.7$~K). Even $m_\text{Cu2}$, although less strongly coupled to the central pair, shows its value significantly renormalized ($\approx \SI{0.7}{\mu_B}$). With this evidence we can establish SeCuO$_3$ as a prime example for further studies of the influence of strong quantum fluctuations on the formation of long-range magnetic order.

The singlet formation at Cu1 sites is also consistent with magnetization measurements \cite{Lee} that show magnetization plateau at half the total magnetization, indicating only Cu2 spin site contribution to total magnetization. Similar dimerisation is reported in spin-tetramer compound CuInVO$_5$ \cite{Hase2016} and in spin system Azurite (Cu$_3$(CO$_3$)$_2$(OH)$_2$) that has two copper sites, with innermost site experiencing only 10\% of fully polarized spin.\cite{Aimo2009}

We would also like to address the fact that the observed exponential dependence of $T_1^{-1}$ is visible down to approximately 60~K below which it starts to saturate. Around the same temperature a simple tetramer model of susceptibility breaks down~\cite{Zivkovic2012}. This shows that some other energy scale starts contributing to $T_1$, most probably the anisotropic DM interaction. The deviation from activated behavior so high in temperature cannot originate from critical fluctuations of AF order, since it onsets only below $T_N =8~$K, and the increase of $1/T_1$ is visible only 1~K above $T_N$. The influence of DM interaction is in agreement with the additional EPR data \cite{Herak2014} where the rotation of the magnetic axis below $\approx 50$~K was ascribed to the DM interaction between Cu1 and Cu2. Lack of Cu2 signal in NQR can then be explained, as Cu2 coupling is weaker and acts as an almost free spin with strong magnetic fluctuations that cause short $T_2$ time. On the other hand, the spin singlet on Cu1 greatly increases the relaxation time thus making that site visible in magnetic resonance experiments. 

In summary, we have determined the magnetic structure of the tetramer compound SeCuO$_3$ by combining neutron diffraction with NQR measurements. We found a highly non-collinear spin configuration with the inner Cu1 spins being antiparallel to each other but forming an angle to the Cu2 spins, indicating an influence of the Dzyaloshinskii-Moriya interaction. Both neutron and NQR measurements detected unexpectedly low ordered moment on the Cu1 site. Combining the spin-lattice relaxation we propose a picture that qualitatively agrees with all the observations. Our results are consistent with strongly-coupled Cu1-Cu1 spins in a tetramer which below $T < J_{11}$ form a spin singlet state. Meanwhile, the Cu2 spins are only weakly coupled to the central pair.

\section{Acknowledgements}
T.C. and M.S.G. are thankful to M. Horvati\'c for suggestions and discussion of NMR results. T.C., M.S.G. and M.P. acknowledge the Croatian Science Foundation (HRZZ) for funding the NMR/NQR studies under Grant No. IP-11-2013-2729. This work has been carried out within the framework of the EUROfusion Consortium and has received funding from the Euratom research and training programme 2014-2018 under grant agreement No 633053. The views and opinions expressed herein do not necessarily reflect those of the European Commission. Neutron diffraction experiments have been carried out at SINQ PSI, Villigen, Switzerland.


\begin{thebibliography}{20}
\bibitem [{mag()}]{magic}%
	\BibitemOpen
	\href@noop {} {}\bibinfo {note} {Magic angle is a root of second order
		Legendre polynomial P{$_2 (\cos \vartheta_m) = 0$}, {$\vartheta_m =
			54.7^\circ$}. At this angle first order quadrupolar interaction vanishes,
		i.e. two inner NQR lines merge.}\BibitemShut {Stop}%
\end{thebibliography}
\end{document}